\documentclass{article}
\usepackage{spconf,amsmath,epsfig}

\usepackage[]{graphicx}
\usepackage{caption}
\usepackage{subfig}
\usepackage{amsmath}
\usepackage{amssymb}
\usepackage{epsfig}
\usepackage{cite}
\usepackage{color}
\usepackage{balance}
\usepackage{amsmath}
\usepackage{amsfonts} 
\usepackage{graphicx}
\usepackage{pdfpages}
\usepackage[T1]{fontenc} 
\usepackage{array}
\usepackage{url}

\usepackage{color}
\usepackage{wrapfig}
\usepackage{lipsum}
\usepackage[hidelinks]{hyperref}
\usepackage{multirow}
\usepackage{tabularx}
\usepackage{gensymb} 

\title{Estimation of solar irradiance\\ using ground-based whole sky imagers}
    
\name{Soumyabrata Dev$^{2}$, Florian M. Savoy$^{1}$, Yee Hui Lee$^{2}$, Stefan Winkler$^{1}$\thanks{This research is funded by the Defence Science and Technology Agency (DSTA), Singapore.}\thanks{Send correspondence to \url{Stefan.Winkler@adsc.com.sg.}}}
\address{
	$^{1}\,$Advanced Digital Sciences Center, University of Illinois at Urbana-Champaign, Singapore \\
	$^{2}$~School of Electrical and Electronic Engineering, Nanyang Technological University, Singapore \\     
}

\begin{document}

\maketitle

\begin{abstract}
Ground-based whole sky imagers (WSIs) can provide localized images of the sky of high temporal and spatial resolution, which permits fine-grained cloud observation. In this paper, we show how images taken by WSIs can be used to estimate solar radiation. Sky cameras are useful here because they provide additional information about cloud movement and coverage, which are otherwise not available from weather station data. Our setup includes ground-based weather stations at the same location as the imagers. We use their measurements to validate our methods.
\end{abstract}


\section{Introduction}
\label{sec:intro}
With the advent of low-cost digital photography, there has been a paradigm shift in the methods used for sensing the earth's atmosphere. Traditionally, images obtained from geo-stationary satellites were the main source of visual information about the earth's atmosphere. These satellite images often suffer from poor temporal and/or spatial resolutions, which makes them less useful for countries like Singapore, where weather phenomena are often highly localized. In these scenarios, high-resolution WSIs can provide more detailed local information of cloud formations. These devices capture the entire sky from the ground at regular intervals with a hemispheric lens. The resulting images are of higher resolution than that obtained from satellites. Also, the upwards pointing nature of the camera allows to capture low-lying clouds.

In our research, we study the effects of clouds and rain on satellite communication links~\cite{cloud_model_compare,IGARSS16-GPS}. We use multiple sensors for this purpose, such as  weather stations, radiosondes and radar. In addition to these meteorological sensors, we also use WSIs providing ample information about clouds. 

In this paper, we explore how a ground-based whole sky imager can be used to estimate the total solar irradiance at its location. We use the luminance of the captured image to estimate solar radiation~\footnote{The source code of all simulations in this paper is available online at \url{https://github.com/Soumyabrata/solar-irradiance-estimation}.}.

The paper is organized as follows: after this introduction, related works are discussed in Section~\ref{sec:rworks}. We describe our setup comprising WSIs and weather stations in Section~\ref{sec:sdata}. Our algorithms and results are presented in Section~\ref{sec:results}. Section~\ref{sec:conclusion} concludes the paper.

\section{Related Work}
\label{sec:rworks}
Huo and Lu~\cite{huo2012comparison} compared cloud coverage obtained from all-sky imager with meteorological observations via field experiments. Similarly, Silva and Souza-Echer \cite{MET1542} compared whole sky imager pictures with human observations for $3$ years of data for two stations in Brazil. Slater et al.\ \cite{slater2001total} compared the cloud fraction derived by two models of sky imagers and their respective algorithms. Furthermore, Yang et al.\ \cite{yang2014solar} as well as Marquez and Coimbra \cite{marquez2013intra} computed short-term forecasts of solar irradiance based on whole sky imagers, using pyranometers for validation.

\section{Data Collection}
\label{sec:sdata}
Our measurements are taken continuously on various rooftops of the Nanyang Technological University (NTU) in Singapore. Our devices comprise three Whole Sky Imagers and three weather stations.  All data collected are sent and archived in a storage and visualization server. 

As a tropical island near the equator, Singapore has a relatively small land mass ($710\mathrm{\ km}^2$) with two monsoon seasons. It witnesses uniform temperatures with rainfall throughout the year.

\subsection{Whole Sky Imagers (WSI)}
Our research group designs custom-made imagers called WAHRSIS (Wide Angle High-Resolution Sky Imaging System)~\cite{WAHRSIS}. We use off-the-shelf components, such as a \emph{Canon} DSLR camera, a \emph{Sigma} fish-eye lens and a single board computer. They are enclosed in a box with a transparent dome for the lens. We have deployed three WAHRSIS \cite{WAHRSIS,IGARSS2015} at NTU, which capture sky/cloud images at intervals of $2$ minutes.

\subsection{Weather Stations}
Weather stations with tipping bucket rain gauges, anemometers and solar sensors are installed on the same rooftops. We use the Davis Instruments $7440$ Weather Vantage Pro II, which collects temperature, humidity, dew point, pressure, surface wind speed and direction, rainfall rate, solar radiation, and solar energy at one minute intervals. 

The  pyranometer provides the total solar radiation reaching a particular area at any instant of time. It is measured in Watt/$\mbox{m}^2$. This solar irradiance is the sum of both direct irradiance component and diffused irradiance component. 

On a clear day with no clouds, the reading of the solar pyranometer follows the clear sky radiation model proposed by Yang et al.\ \cite{dazhi2012estimation}.
The clear-sky Global Horizontal Irradiance (GHI) model for Singapore is given by:
\begin{align}
\label{eq:GHI-model}
G_c = 0.8277E_{0}I_{sc}(\cos\theta_z)^{1.3644}e^{-0.0013\times(90-\theta_z)},
\end{align}
where $G_c$ is the clear-sky GHI in W/$\mbox{m}^2$, $E_{0}$ is the eccentricity correction factor of earth, $I_{sc}$ is a solar irradiance constant ($1366.1W/m^2$), $\theta_z$ is the solar zenith angle (in degrees). The correction factor $E_{0}$ is computed as follows:
\begin{equation*}
\begin{aligned}
\label{eq:E0value}
E_0 = 1.00011 + 0.034221\cos(\Gamma) + 0.001280\sin(\Gamma) + \\0.000719\cos(2\Gamma) + 0.000077\sin(2\Gamma).
\end{aligned}
\end{equation*}
$\Gamma = 2\pi(d_n-1)/365$ is the day angle (in radians), where $d_n$ is the day number in a year.

\begin{figure}[htb]
\hspace{-5mm}\includegraphics[width=1.15\columnwidth]{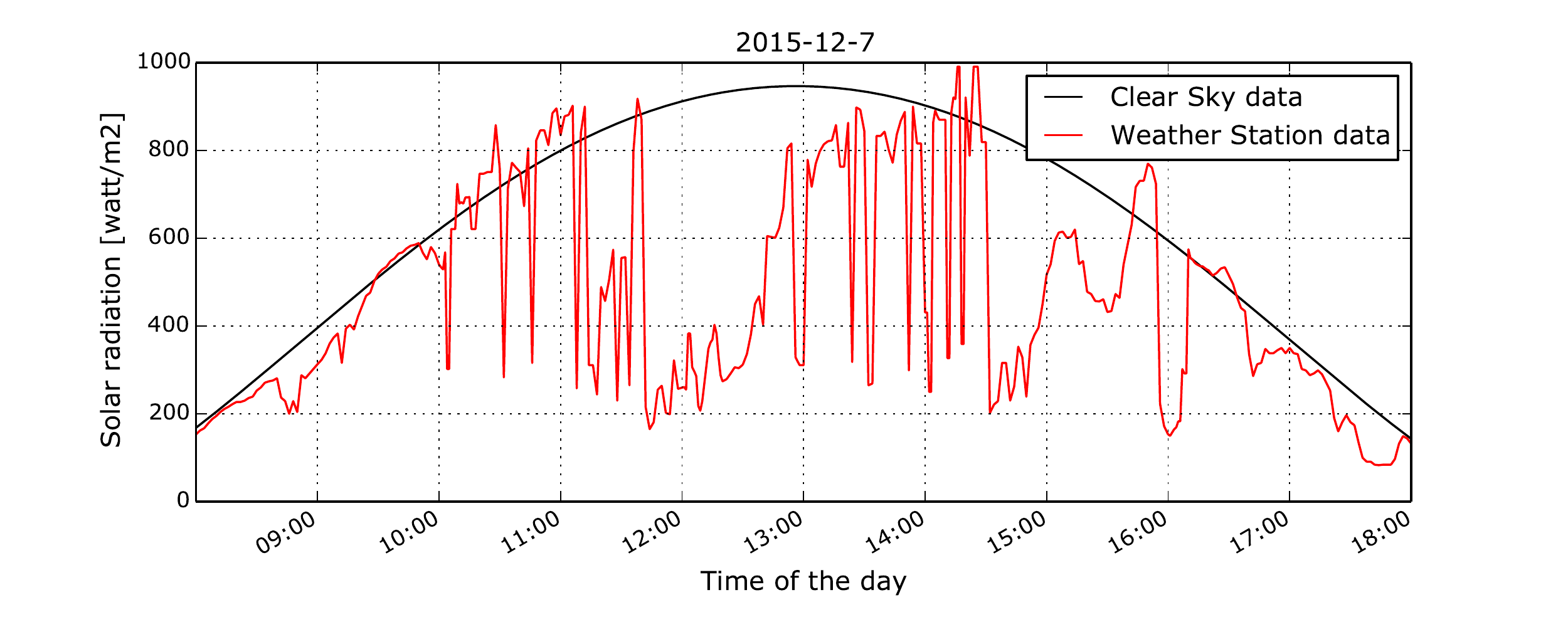}
\caption{Clear-sky solar radiation using \cite{dazhi2012estimation} and actual solar radiation measured at NTU Singapore on 7 December, 2015.
\label{fig:typical_day}}
\end{figure}

We illustrate this relationship by showing an example of clear-sky radiation and actual solar radiation measured on 7 December 2015 on our university campus in Fig.~\ref{fig:typical_day}. We observe that the measured solar radiation fluctuates rapidly as compared to the clear-sky radiation. 

\section{Algorithms and Experimental Results}
\label{sec:results}
In this section, we discuss our experimental results on the estimation of solar irradiance from WSI images. We also discuss how clouds around the sun have a higher impact on the total solar radiation, as compared to the clouds near the horizon.

\subsection{Direct Solar Radiation}
The pyranometer reads the instantaneous solar irradiance reaching the sensor at any instant of time. This constitutes both the direct and diffuse component of solar radiation. On a typical day, the solar radiation can fluctuate significantly in a short span of time, which is quite different from the ideal cosine response. If we examine the  solar radiation readings on the $7$th of December $2015$ from  (cf.\ Fig.~\ref{fig:typical_day}), the recorded solar radiation falls drastically from 758 Watt/$\mbox{m}^2$ at 10:30, to 283 Watt/$\mbox{m}^2$ at 10:32, and then rises again to 714 Watt/$\mbox{m}^2$ at 10:34.
We observe two rapid fluctuations within a span of a few minutes.

\begin{figure}[htb]
\begin{center}
\includegraphics[height=0.14\textwidth]{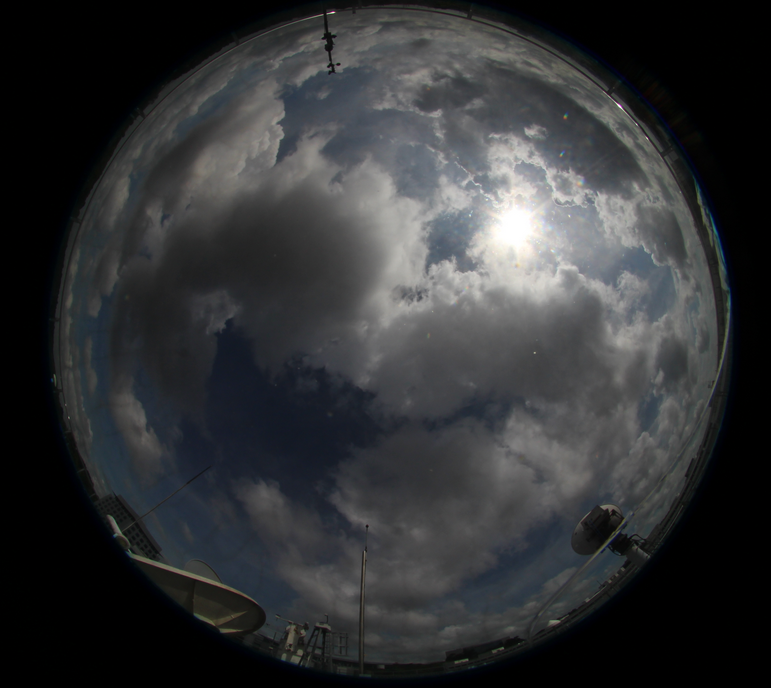}\hspace{-1mm}
\includegraphics[height=0.14\textwidth]{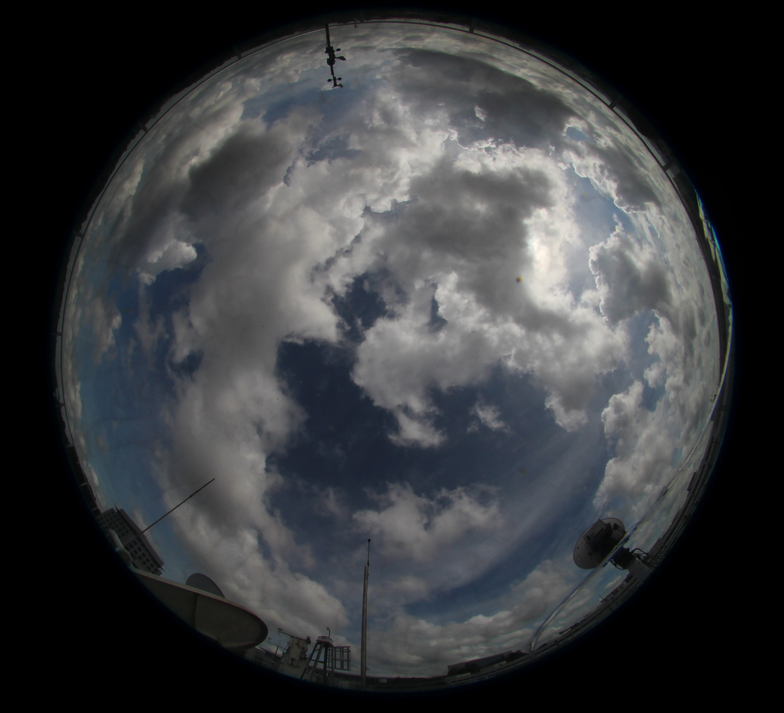}\hspace{-1mm}
\includegraphics[height=0.14\textwidth]{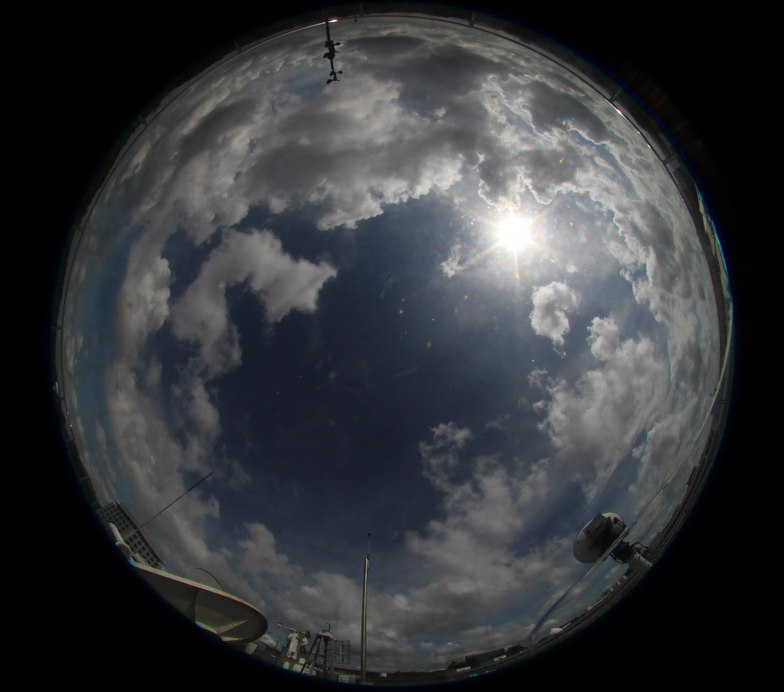}\\
\makebox[0.15\textwidth][c]{\small 10:30 (758 W/$\mbox{m}^2$)}
\makebox[0.15\textwidth][c]{\small 10:32 (283 W/$\mbox{m}^2$)}
\makebox[0.15\textwidth][c]{\small 10:34 (714 W/$\mbox{m}^2$)}
\caption{Consecutive images captured over an interval of $4$ minutes on $7$ December, $2015$. The amount of total solar radiation recorded by the weather station is shown in parentheses. The sun is obscured by clouds at 10:32, resulting in a sudden dip in solar radiation.
\label{fig:fluctuation_example}}
\end{center}
\end{figure}

The images captured using WAHRSIS at the corresponding time instants are shown in Fig.~\ref{fig:fluctuation_example}.  It becomes clear  that the fluctuation in solar radiation is due to  clouds obscuring the sun from view.  Assuming the diffuse component of solar radiation to be almost constant at these three instants of the day, the fluctuations in solar radiation recording are primarily because of the direct irradiance component. Therefore, the clouds in the vicinity of sun have the highest impact on recorded solar radiation.

\subsection{Estimating Solar Irradiance}
\begin{figure*}[htb]
\begin{center}
\includegraphics[width=0.95\textwidth]{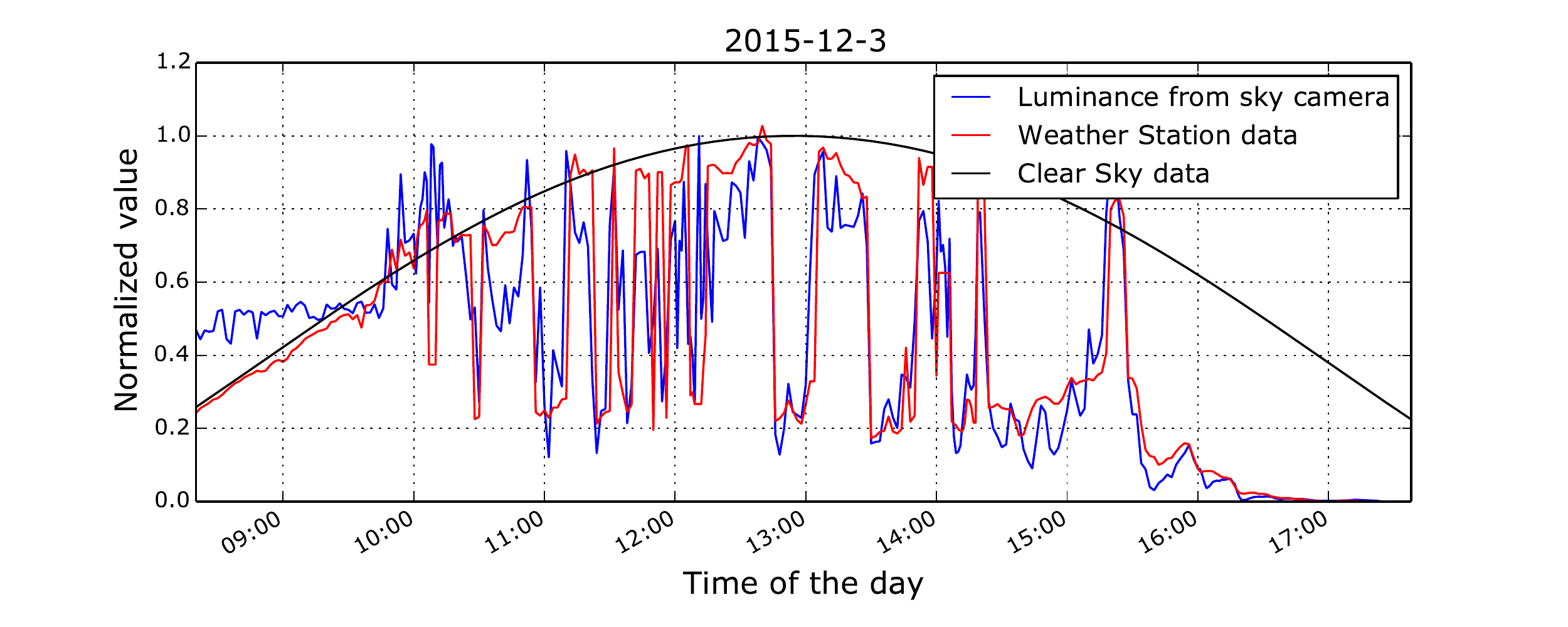}
\caption{Solar radiation measured by the weather station (in red) and computed by our proposed method (in blue) for $3$rd December 2015. The black curve is the clear sky solar radiation model presented from \cite{dazhi2012estimation}.}
\label{fig:lum_rad}
\end{center}
\end{figure*}

In this section, we present our algorithm to estimate solar radiation from WSI images. We crop a square patch of a fixed size around the sun from the image. The position of the sun in image is detected by setting a threshold of $240$ in the red channel of the image~\cite{IGARSS2016a}. The centroid of the largest generated polygon from thresholding is considered as the sun's position in image. We thereby compute the mean luminance of the cropped image by averaging the luminance channel of the image in the HSL (Hue, Saturation, Lightness/Luminance) color space.

The capture settings might differ from one image to another. In our setting, the aperture and ISO are kept constant, and only the exposure time  varies. We thus divide the computed luminance by the exposure time. The resulting number can be interpreted as the relative luminance that the camera would have captured if the shutter was open during a second, neglecting over-saturation effects. We finally normalize this value with respect to the clear sky radiation~\cite{dazhi2012estimation} across the entire day.

\begin{figure}[htb]
\begin{center}
\includegraphics[width=0.5\textwidth]{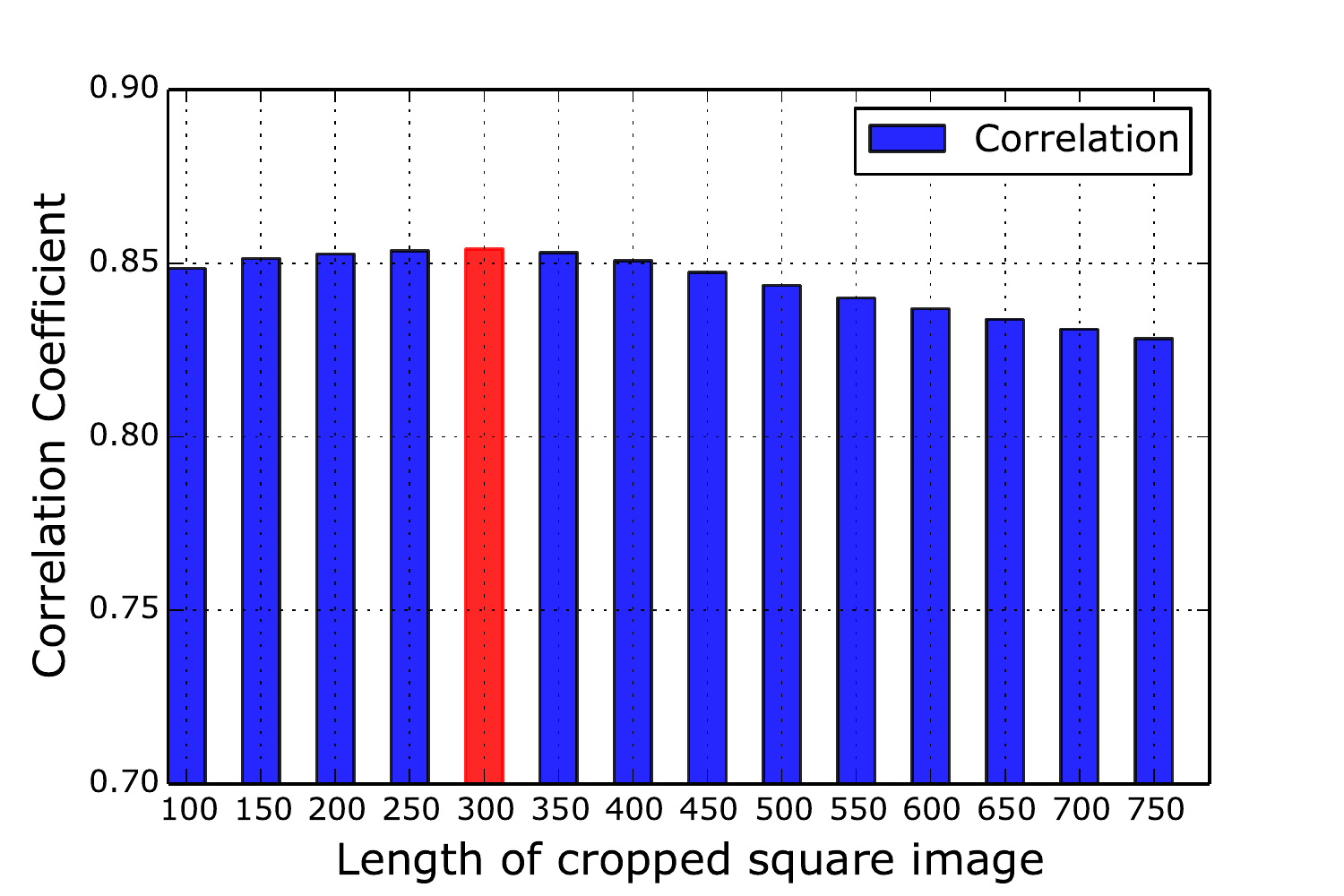}
\caption{Impact of image patch size on correlation. Best performance is obtained for a crop size of $300 \times 300$ (red bar).
\label{fig:cropsize}}
\end{center}
\end{figure}

In order to fix the cropped patch size, we measure the correlation  between solar radiation and luminance for different crop sizes around the sun.  We perform experiments using $563$ images on a typical day of December $2015$, and show the results in Fig.~\ref{fig:cropsize}. We obtain the optimal performance for a size of $300 \times 300$, which we select for the subsequent experiments.

\begin{figure}[htb]
\begin{center}
\includegraphics[width=0.9\columnwidth]{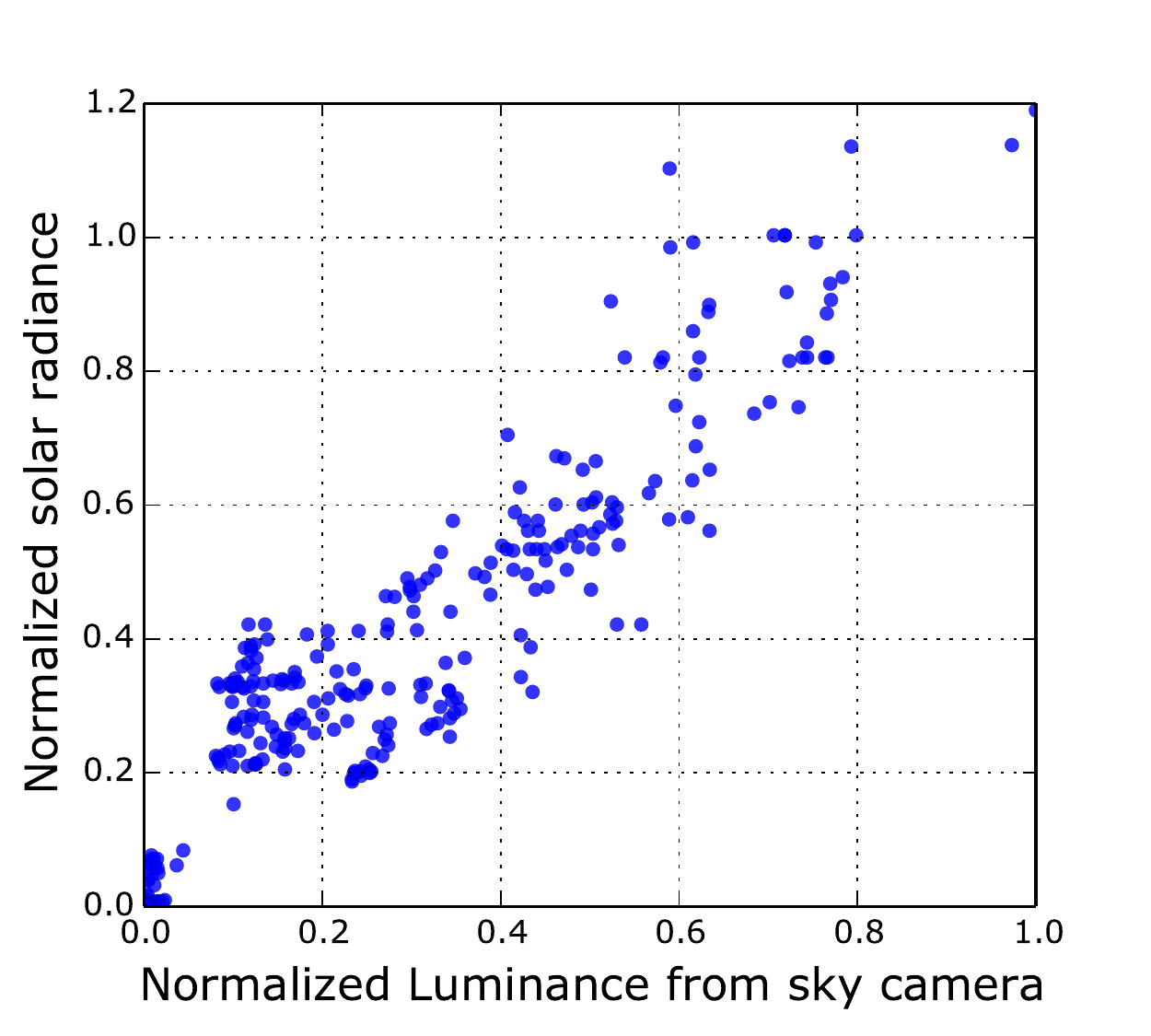}
\caption{Scatter plot between solar irradiance and luminance for the month of December 2015.}
\label{fig:scatter_Dec2015}
\end{center}
\end{figure}

Figure~\ref{fig:lum_rad} shows the measurements for the $3$rd of December $2015$, with the weather station data in red and the estimated solar radiation in blue. We see that both strongly follow each other. The study was performed in December during the Northeast Monsoon, with rainfalls usually occurring in afternoons and early evenings. This relates with lower solar radiation towards the end of the day.

We test our proposed algorithm for a large period of time. We perform this experiment for all days in the month of December 2015. Figure~\ref{fig:scatter_Dec2015} shows the scatter plot between normalized solar irradiance value and normalized image luminance in the vicinity of the sun. We observe a strong correlation (Pearson correlation coefficient of $0.93$).  

\section{Conclusions}
\label{sec:conclusion}
In this paper, we have shown how sky images can be used to estimate solar radiation. Our method focuses on the direct solar radiation as computed from the image luminance in the vicinity of the sun. We used one month of weather measurements taken at the same location to validate our proposed algorithm.   

In future work, we will use other information such as cloud coverage~\cite{ICIP2015a} and cloud type recognition~\cite{ICIP2015b} obtained from sky images for more accurate estimates and now-casting of solar radiation.  We also plan to investigate the effects of rain events on image luminance. 

\balance

\bibliographystyle{IEEEbib}

\end{document}